\documentclass[conference]{IEEEtran}
\usepackage{blindtext, graphicx, widetext}

\ifCLASSINFOpdf
\else
\fi
\begin{document}
%
\title{Familiar Strangers: the Collective Regularity in Human Behaviors}

\author{\IEEEauthorblockN{Yan Leng}
\IEEEauthorblockA{MIT Media Lab\\
Cambridge, MA, USA\\ Email: yleng@mit.edu}
\and
\IEEEauthorblockN{Dominiquo Santistevan}
\IEEEauthorblockA{Electrical Engineering \& Computer Science\\
Cambridge, MA, USA\\
Email: niquo@mit.edu}
\and
\IEEEauthorblockN{Alex 'Sandy' Pentland}
\IEEEauthorblockA{MIT Media Lab\\
Cambridge, MA, USA \\
Email: pentland@mit.edu}}


\maketitle


\begin{abstract}
    The social phenomenon of familiar strangers was identified by Stanley Milgram in 1972 with a small-scale experiment. However, there has been limited research focusing on uncovering the phenomenon at a societal scale and simultaneously investigating the social relationships between familiar strangers. With the help of the large-scale mobile phone records, we empirically show the existence of the relationship in the country of Andorra. Built upon the temporal and spatial distributions, we investigate the mechanisms, especially collective temporal regularity and spatial structure that trigger this phenomenon. Moreover, we explore the relationship between social distances on the communication network and the number of encounters and show that larger number of encounters indicates shorter social distances in a social network. The understanding of the physical encounter network could have important implications to understand the phenomena such as epidemics spreading and information diffusion.     
\end{abstract}

It is a common phenomenon to encounter strangers during regular activities in our daily lives whom we can recognize but never formally interact with \cite{milgram1992individual, paulos2004familiar}. This intriguing social phenomenon of "Familiar Stranger" of the urban environment, first identified by Stanley Milgram in 1972 by asking travelers to recognize ''strangers'' they met on a bus platform \cite{milgram1992individual}. In 2013, Sun \cite{sun2013understanding}, for the first time, uncovered the encounter mechanisms of three million public transit users in Singapore to capture the time-resolved, in-vehicle encounter patterns and familiar strangers using transit smart-card data.

This hidden dynamic social network plays an unnoticeable but significant role in information diffusion, behavior synchronization, and epidemics spreading process. Christakis \cite{christakis2013social} and Montanari \cite{montanari2010spread} shed light on the diffusion of information, innovations and behaviors via social contagion driven by social interactions. Dong used a Markov jump process to capture the co-evolution of friendship and visitation patterns in a student dorms with monthly surveys and locations tracking through mobile phones \cite{dong2011modeling}. Besides, a series of studies have focused on using large-scale or high-resolution data to empirically and computationally model the epidemics process. Danon showed that large-scale interaction data are needed to verify the assumptions of random transmission models and simple network structures to understand better and predict the disease transmission \cite{danon2012social}. Stopczynski utilized large-scale behavioral data for modeling epidemic prediction by physical proximity network \cite{stopczynski2015physical}. Isella and Salathe tracked high-resolution proximity network to understand the transmission paths of diseases \cite{isella2011s, salathe2010high}. 

To capture the physical proximity network in the urban environment, we use countrywide mobile phone logs to explore the phenomenon and underlying mechanisms that trigger familiar strangers on a societal scale covering many aspects of social lives. This data simultaneously captures two layers of a network: physical proximity and social networks. To our knowledge, we are the first to identify country-wide familiar strangers utilizing both mobility and social networks. We confirm the existence of the phenomenon of a familiar stranger in an urban environment. At a macro scale, we found that the collective regularities - temporal regularity and spatial structure - explain the phenomenon of familiar strangers. We also investigate the relationship between physical co-occurrences and the proximity in social networks and show that physical co-occurrences indicate shorter social distances via a communication network. A series of landmark papers have established the regularity and predictability of human mobility \cite{gonzalez2008understanding, song2010limits}.  Moreover, the understanding of collective regularity has implications for the prevention of epidemics and the facilitation of information spreading. 

We organize the paper as follows. We first describe the data set and the settings we use to conduct our study. We test our hypothesis on the mechanism that drives the collective behavior of encountering. In the end, we overlay the physical proximity network with phone communication network to explore the relationships between them.

\section*{Data and settings}
\label{data}

To capture both the social network and mobility network, we study the anonymized Call Detail Records of Andorra, a European country, for July 2016. This dataset includes the caller, receiver, connected cell tower, start and end time of the connection. The spatial and social network encoded in Call Detail Records enable us to identify familiar strangers. We created a communication network where each user is a vertex, and an edge exists if there exists direct contact between two users. In July, we are looking at a total of 1,264,292 users. After filtering out users with more than 100 connections that may be hotels or vendors, we only consider the remaining 1,211,814 users. We identify the physical encounters by observing if two “users” called or text on the same tower within a one-hour window of each other. In our study, we define familiar strangers as individual pairs whom physical co-locate at one cell tower within the one-hour time window, but there exist no direct links on the communication network. 

\section*{Spatial and temporal patterns of encounters}

\begin{figure}[tbhp]
\centering
\includegraphics[width=\linewidth]{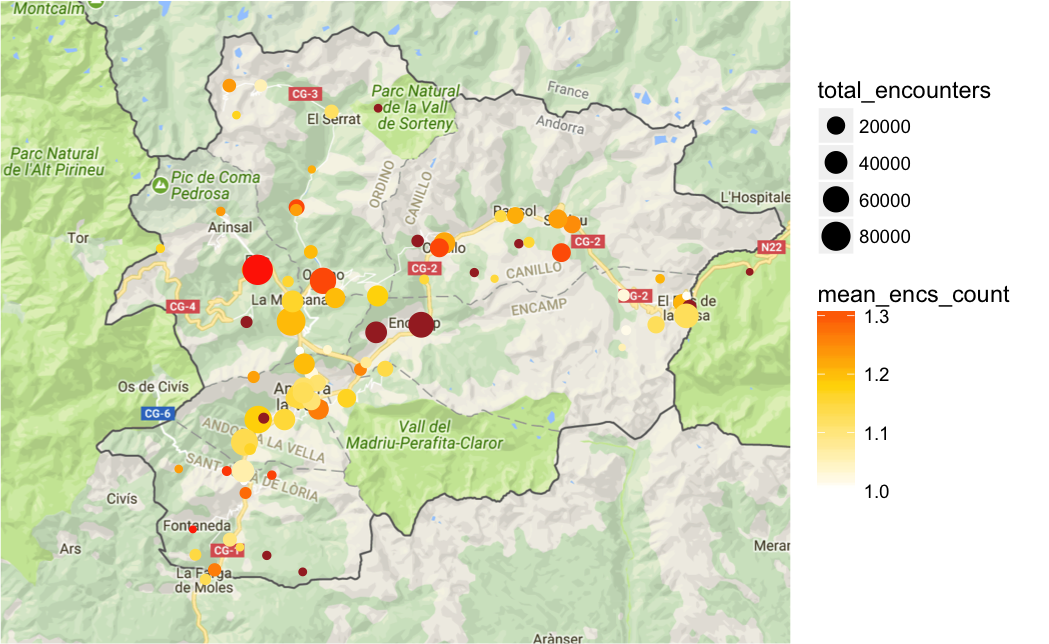}
\caption{Map of cell towers, where the size of the node is proportional to total number of encounters and the color is darker where a pair of users encounter more times on that tower.}
\label{fig:spatial_encs_map}
\end{figure}

To understand the temporal distribution of physical encounters, we extract the time when an encounter happens as we see in Figure \ref{fig:spatial_encs_map}. There existed prominent spikes between working hours (8 am - 11 am and 3 pm - 5 pm) on weekdays, and slight shifts for weekends (10 a.m. - 11 a.m. and 3 p.m. - 6 p.m.). Interesting, we see a small peak at 11 p.m. on Saturday night, which captures the encounters of Saturday nightlife. 

\begin{figure}[tbhp]
\centering
\includegraphics[width=\linewidth]{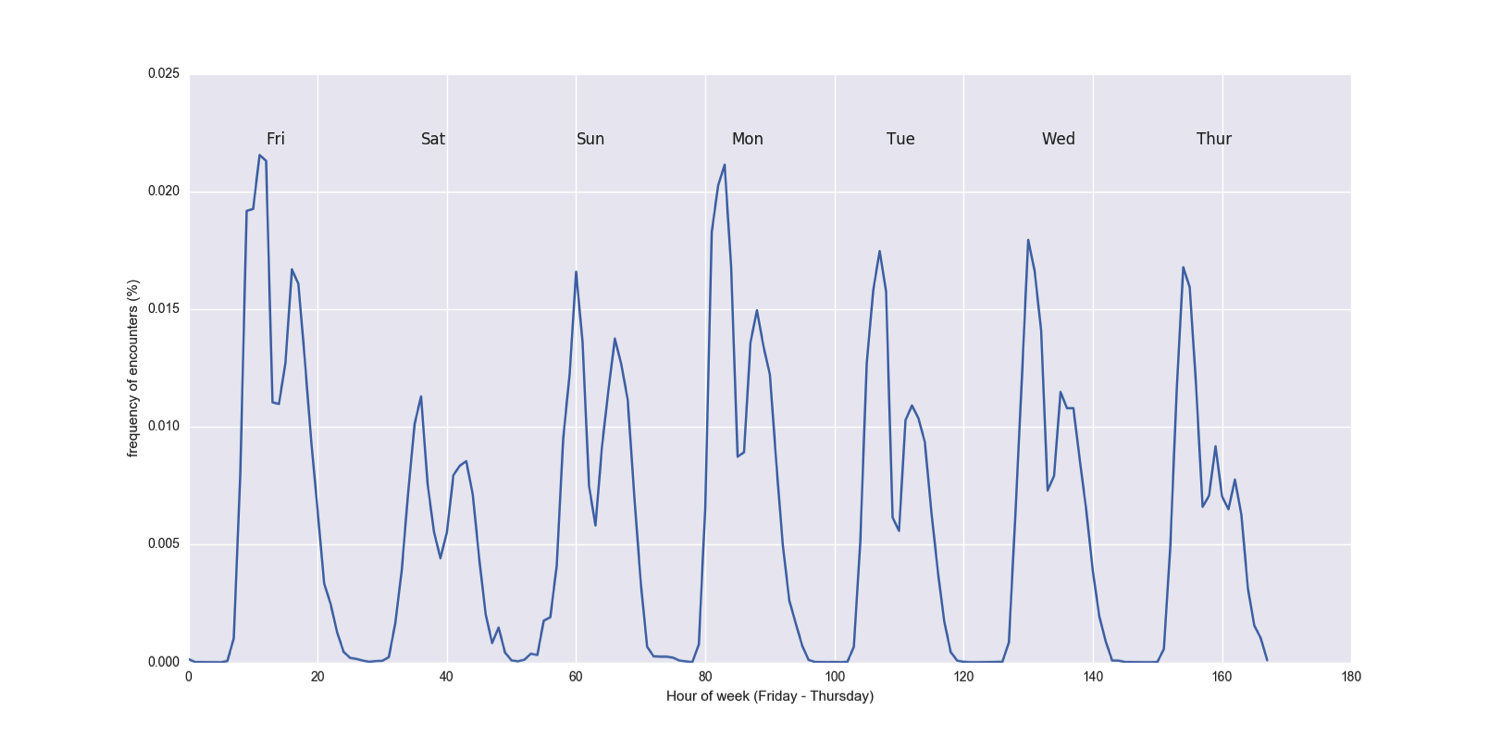}
\caption{Temporal distribution of physical encounters}
\label{fig:temporal_pattern_encs}
\end{figure}

We also analyze the spatial distribution of encounters, and the number of encounters happens for each encounter pair. More specifically, each tower has different usage that is dependent on the popularity around that tower, but there is also a distribution over how many encounters happen between two users at a single tower. As shown in Figure \ref{fig:spatial_encs_map}, there exists some towers that are not quite as popular with encounters, yet they have some of the highest numbers of encounters counts between pairs of users. 

\section*{Collective regularities}
\label{mechm}

Human mobility follows a high degree of temporal and spatial regularity \cite{gonzalez2008understanding}, and physical encounter does not happen out of ``coincidence''. We believe that collective regularity of human behaviors triggers the physical co-occurrences. In particular, we focus on the temporal regularities and spatial structure of two consecutive encounters to uncover the underlying mechanisms that drive the encounter.

\subsection{Temporal regularity}
Human daily routines, such as commuting, follow certain temporal patterns. To explore the repeated encounters at a population scale, we create an encounter network based on mobility behaviors across a week and measure the inter-event time $\Delta t$ between consecutive encounters of each familiar stranger pair. 

\begin{figure}[tbhp]
\centering
\includegraphics[width=.9\linewidth]{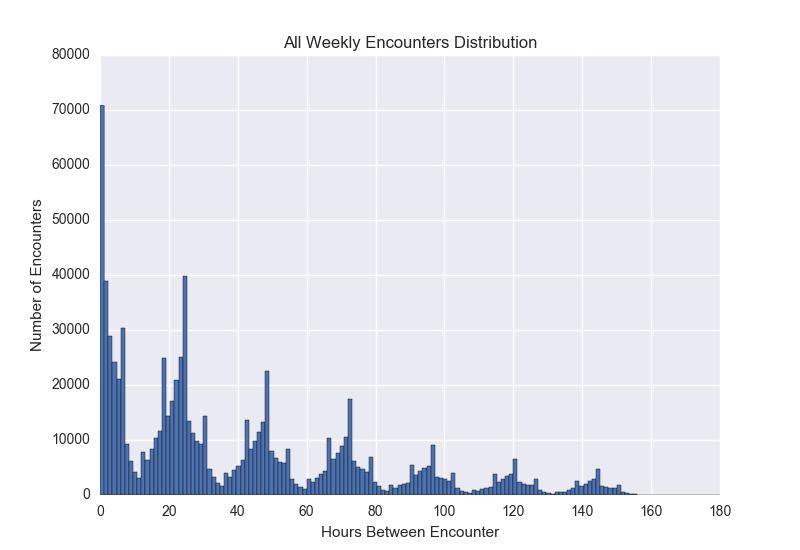}
\includegraphics[width=.8\linewidth]{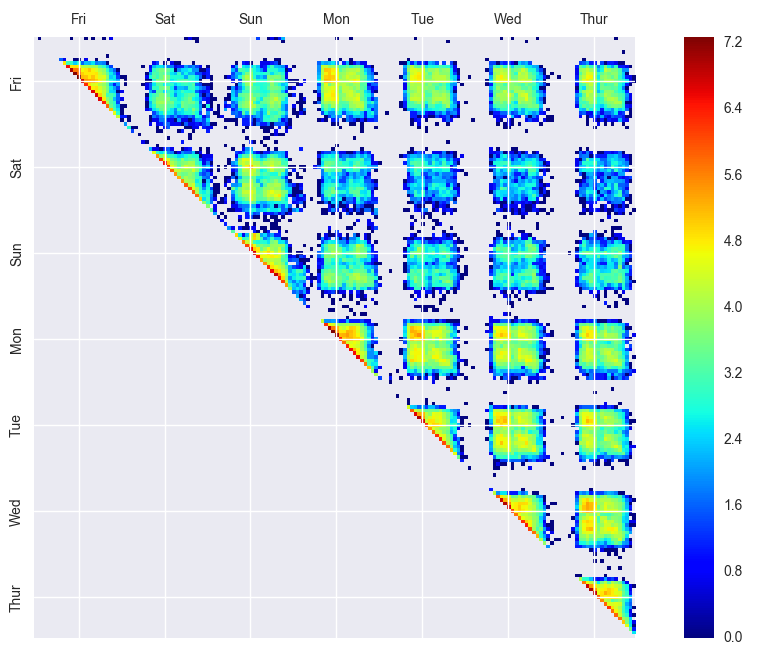}
\caption{Temporal regularity of collective mobility behaviors. The left panel shows the probability density function of inter-event time ($P_{\Delta t}$ between two consecutive encounters. The right panel shows the encounter time of two consecutive co-locations}
\label{fig:enc_reenc_dist}
\end{figure}

The left panel of Figure \ref{fig:enc_reenc_dist} shows the distribution of inter-event time between two encounters. We observe prominent peaks for every 24 hours and another lower peaks for $24 \cdot d \pm 6$ hour, where $d$ is the $d^{th}$ day after the first encounter. This pattern indicates that people are very likely to encounter their familiar strangers on the same hour of a day within the next $d$ days. Also, we analyze the time of the consecutive encounters. We categorize the right panel of Figure \ref{fig:enc_reenc_dist} into two types of encounters - weekday encounters and weekend encounters. Specifically, people who encounter during weekdays are less likely to encounter each other during weekends and vice versa. Besides, morning encounters are more likely to encounter again during the morning, explaining the 24-hour peak as shown in the figure in the left panel. Both empirical exercises highlight the collective temporal regularity in people's daily routines partly explains the repeated physical encounters. 

\subsection{Spatial structure}

We also investigate the spatial structure of repeated encounters, specifically the popularity of the two locations, distances between the two locations and the Points of Interests. 

\subsubsection{Popularity of and distances between the sequential encounter locations} 
\begin{figure}[tbhp]
\centering
\includegraphics[width=.8\linewidth]{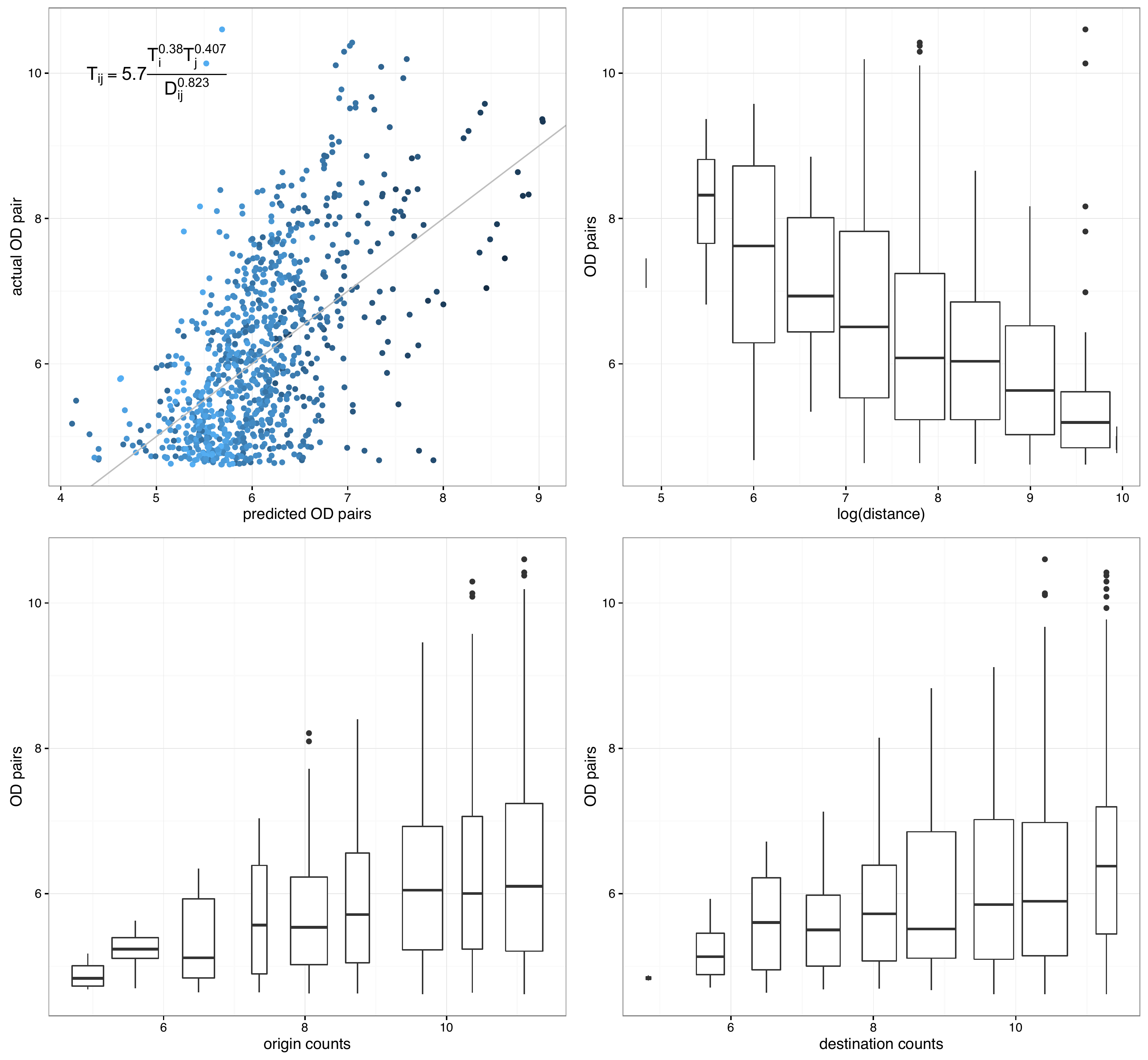}
\caption{Gravity law fit of encounter network. (A) Encounter and re-encounter flow obtained from data on the number of encounters the first and second locations, as a function of the distance between the encounter and re-encounter locations. (B-D) Encounter and re-encounter flow as a function of distance, encounter popularity of first and second locations.}
\label{fig:gravity}
\end{figure}

To gain insights on the encounter and re-encounter pairs, we use the general gravity model to understand the relationships between spatial distances, attractions (popularity) of the two locations. The gravity model utilized in our study assumes the following functional form, as shown in (\ref{eq:gravity}). We demonstrate the fit of the model and the relationships between encounter/ re-encounter flow and the variables of interests in Figure \ref{fig:gravity}. 
\begin{equation}
    T_{ij} = C \frac{N_i^\alpha  N_j^\beta}{D_{ij}^\gamma}
    \label{eq:gravity}
\end{equation}

where $T_{ij}$ is the number of encounter and re-encounter flow between two geographical location $i$ and $j$. $D_{ij}$ is the distance between two geographical areas and $N_i$ and $N_j$ are the number of encounters of area $i$ and $j$ respectively. After applying a logarithmic transformation and fit parameters with a linear regression, we found $\alpha = 0.38, \beta = 0.407, \gamma = 0.823$. 

\subsubsection{Points of Interests}  

\begin{figure}
\centering
\includegraphics[width=.8\linewidth]{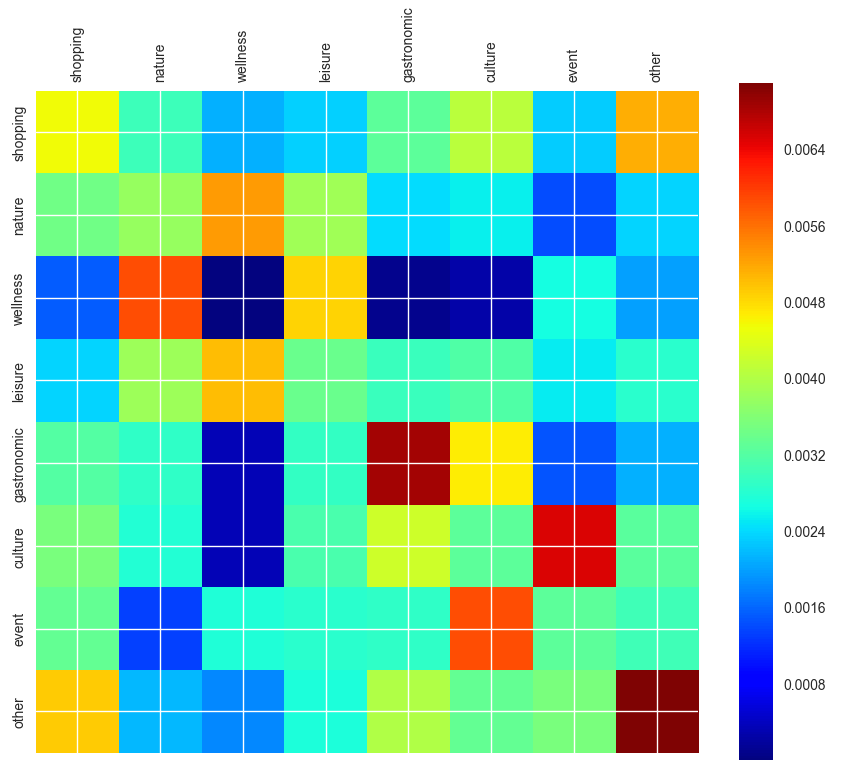}
\caption{Normalized transition probabilities from one to another type of POIs. X-axis is the POI type of the first encounter location and Y-axis is the POI type of the re-encounter location}. 
\label{fig:activity}
\end{figure}

The analysis of the Points of Interests (POIs) surrounding cell towers enables us to infer trip purposes, the routines and the interests of the individual (\cite{leng2017synergistic,leng2016urban}). By analyzing the Points of Interests attached to the two consecutive locations, we have some intriguing empirical findings and need further investigation and generalization. As shown in Figure \ref{fig:activity}, the warmer the color, the more likely encounter and re-encounter happen at type A and type B POIs. As indicated by the color of each cell, the high probability encounter and re-encounter pair are food - food, the stadium - culture, culture - event and wellness - nature. The preceding pattern indicates that people who share the same interests or within the same socio-demographic group are likely to meet each other at another type of locations, which is in line with Milgram's finding in a small-scale experiment. Crandall (2010), has the similar observation on a large-scale data, in particular, a minimal number of co-occurrences results in a high likelihood of a social tie, the negation of which may induce noises in masking acquaintances to be strangers \cite{crandall2010inferring}.  


\section*{Familiar strangers in social network}

\begin{figure}[tbhp]
\centering
\includegraphics[width=\linewidth]{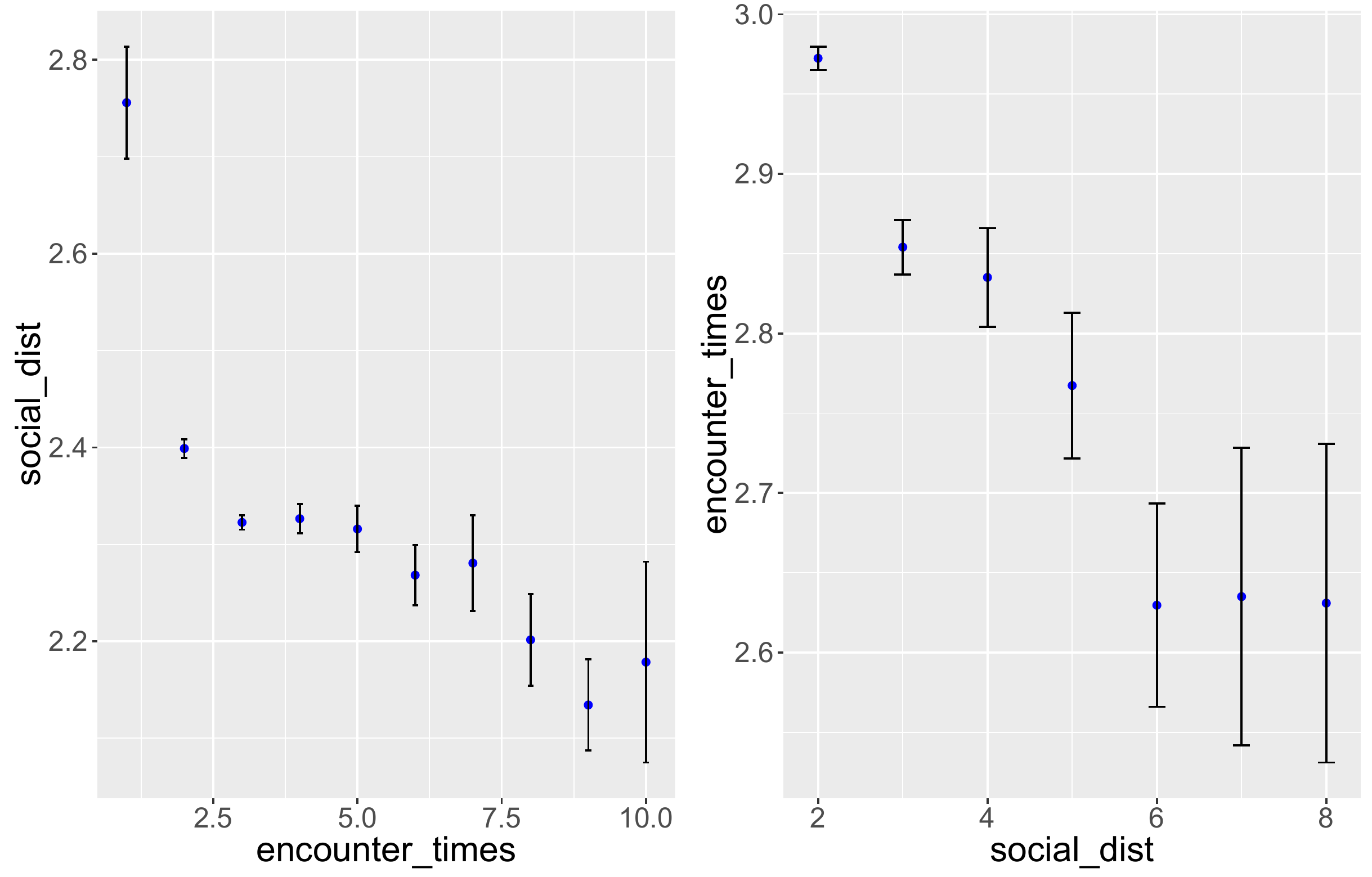}
\caption{Relationship between number of encounters and social distances in social networks of familiar stranger pairs}
\label{fig:social_dist_encounter_times}
\end{figure}

\begin{figure}[tbhp]
\centering
\includegraphics[width=.8\linewidth]{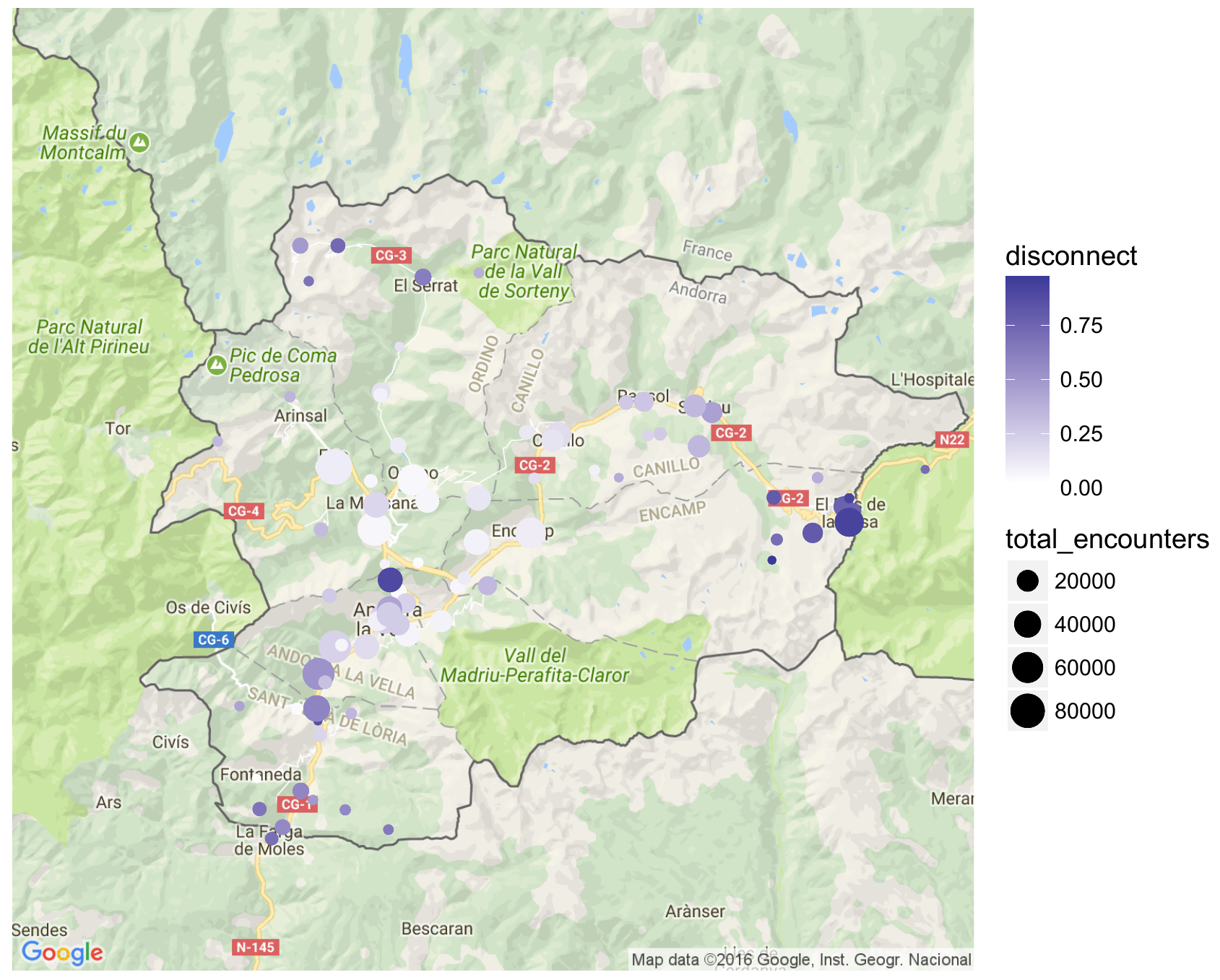}
\includegraphics[width=.8\linewidth]{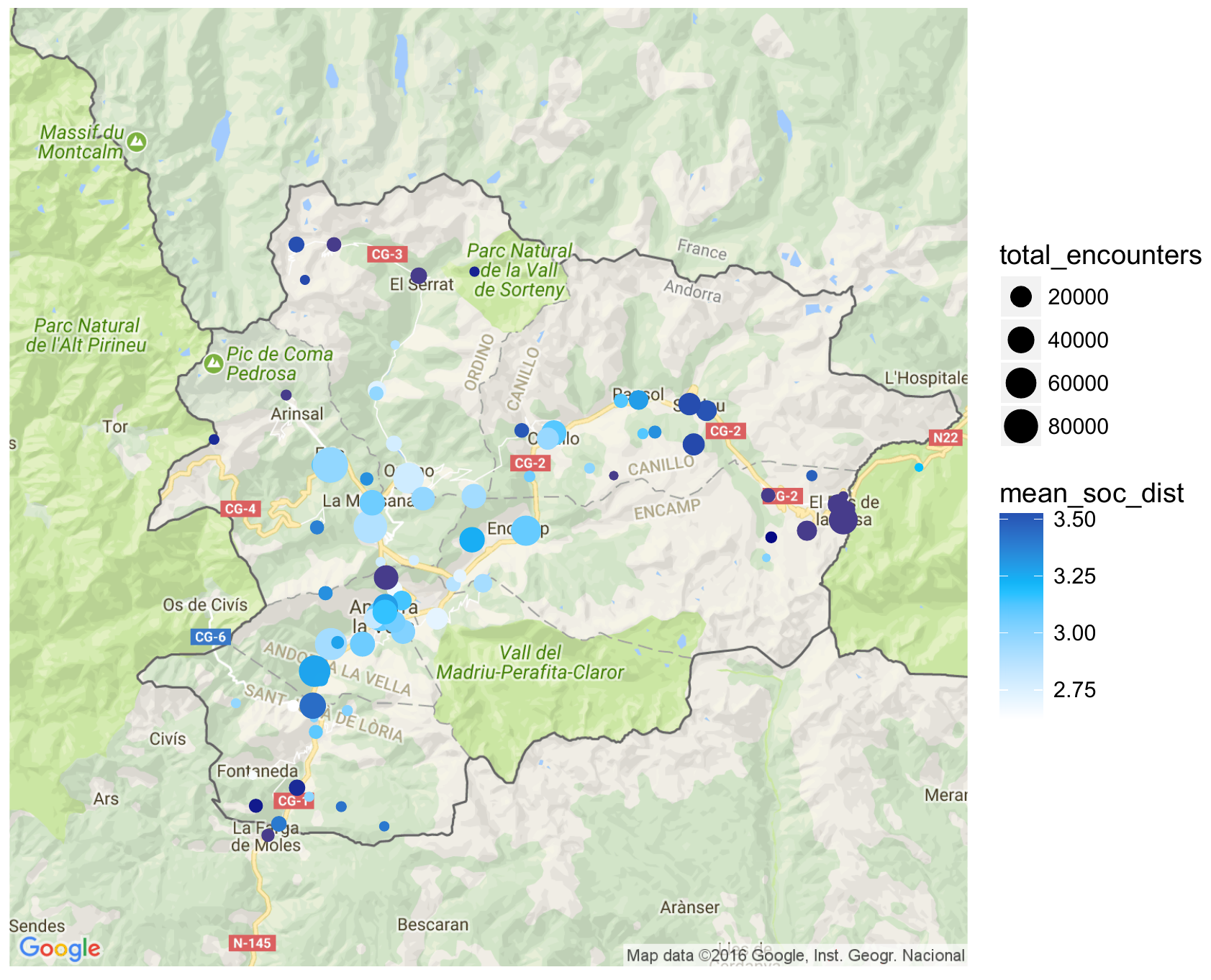}
\caption{Spatial distributions of disconnected individuals in social network and social distances  between two individuals in encounter pairs}

\label{fig:encouner_social}
\end{figure}

There exist studies establishing the relationships between mobility behaviors and social ties. Crandall (2010) developed a framework to empirically and mathematically investigate the relationship between social ties and co-occurrence \cite{crandall2010inferring}. Along the same line, Toole (2015) found that that the composition of a user's ego network concerning the type of contacts they keep correlates with mobility behavior\cite{toole2015coupling}. Apart from the observed relationship between mobility behavior and the probability of the formation of a tie or the strength of a tie, we are specifically interested in the social distances between 'strangers' who physically encounter one another multiple due to similar routines or interests. As shown in Figure \ref{fig:social_dist_encounter_times}, there exists a negative relationship between social distance and the number of encounters - the more time each familiar stranger pair encounter one another, the closer they are on the social networks. The spatial distributions of disconnected individuals and social distances in the social networks are shown in Figure \ref{fig:encouner_social}. We observe that the closer the encountering to the center of the country, the less likely two individuals will be disconnected and closer they are on the social network.

\subsection{Formation of community}

\begin{figure}[tbhp]
\centering
\includegraphics[width=.8\linewidth]{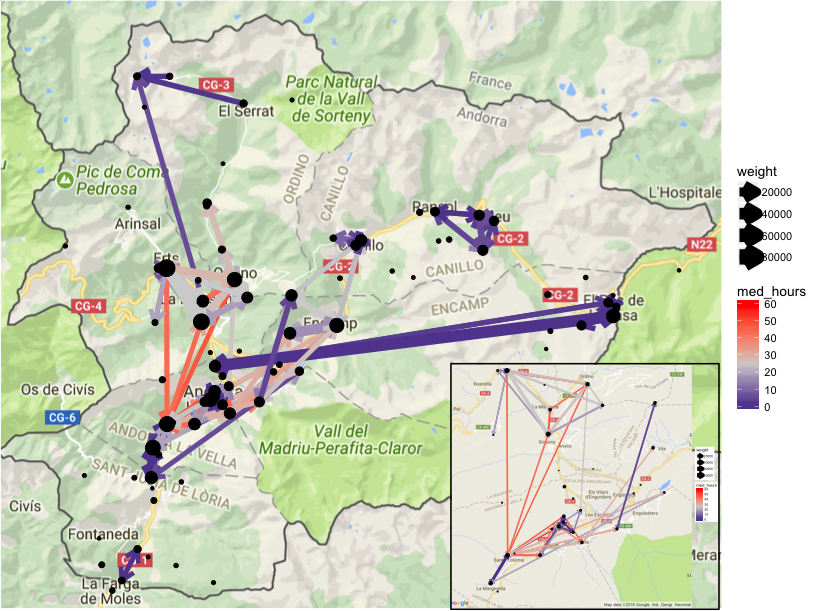}
\caption{Time-resolved encounter-reencounter network. Links with less than 2000 encounter-reencounter pairs within a week is removed. The width is the links represents the counts of encounter-reencounter paris and the color of the links represents the time interval between encounter and reencounter.}
\label{fig:community}
\end{figure}

As stated by Milgram, it is easier for familiar strangers to talk to each other if they meet in a different location. In other words, collective behavior caused by similar inherent routines may form an actual friendship due to a spatiotemporal coincidence. In Figure \ref{fig:community}, we show a time-resolved encounter-reencounter network. We can see that there exist communities connected by nearby cell towers, such as Canillo, La Massana, and Encamp. However, the areas linked by warmer color are interesting in that long-interval encounters to link them, which indicates similar daily schedules. Pairs colored in purple note similar trajectory within 12 hours. Grey colors indicate people may meet around 24 hours and Red red color indicate re-coincidence at around two days. 


\section*{Discussions}

The phenomenon of a familiar stranger has been formally identified by a small-scale experiment by Stanley Milgram since 1972. However, there has been limited studies focusing on proving the phenomenon at a societal scale and investigating the social relationships between familiar strangers. With the help of the large-scale mobile phone records, we empirically show the existence of the link in the country of Andorra. Built upon the temporal and spatial distributions, we investigate the mechanisms that trigger this. In the end, we explore the relationship between social distances on the communication network and the number of encounters and show that more substantial amount of encounters indicate shorter social distances on the social network.

In this study, we use the large-scale Call Detail Records of a European country Andorra to create the physical encounter network and phone communication network. We show the existence of the familiar stranger phenomenon in an urban environment. By analyzing the temporal and spatial characteristics of the encounters, we uncover the underlying mechanisms, especially collective temporal regularity and spatial structure that trigger the phenomenon. In the end, we explore the relationship between social distances along social network and number of encounter in mobility network. We show that a more substantial amount of encounters predicts shorter social distances on social networks. The understanding of the physical encounter network could have significant implications for epidemics preventions and information spreading facilitation. 

Our study posits several interesting future studies. First of all, one natural future work is to investigate how the relationship of familiar strangers grows into an actual friendship, and how their behaviors intersect and synchronize with each other. It would be even more interesting to understand the causal relationships between physical co-occurrences and the formation of social ties. Another promising direction is to integrate the familiar stranger relationships into the modeling and simulation of epidemics spreading and information diffusion.

\bibliography{main}
\bibliographystyle{IEEEtran}

\end{document}